\documentclass[10p,times,twocolumn]{article}

\usepackage{geometry}
\usepackage{amssymb}

\usepackage{amsmath}
\usepackage{subfigure} 
\usepackage[T1]{fontenc}

\usepackage{ marvosym }
\usepackage{tikz,pgfplots} 

\usepackage[square,sort,comma,numbers]{natbib}
\usepackage[affil-it]{authblk}
 \usepackage { amsmath }

\begin{document}

\title{\textbf{Translational and rotational dynamics of a large~buoyant~sphere in turbulence}}


%
%
%
%
%
%
%

\author{Varghese Mathai$^{1}$ \thanks{\Email: v.mathai@utwente.nl }}
\author{Matthijs W. M. Neut$^{1}$}
\author{Erwin P. van der Poel$^{1}$}
\author{Chao Sun$^{2,1}$ \thanks{\Email: chaosun@tsinghua.edu.cn}}
\affil{\normalsize{ $^{\textup{1}}$ Physics~of~Fluids~group, Mesa+~Institute,  and J.\ M.\ Burgers~Centre~for~Fluid~Dynamics, University~of~Twente, 7500 AE Enschede, The Netherlands}}
\affil{\normalsize{$^{\textup{2}}$ Center for Combustion Energy and Department of Thermal Engineering, Tsinghua University, 100084 Beijing, China}}

\maketitle

\paragraph{Abstract:} \textcolor{black}{We report experimental measurements of the translational and rotational dynamics of a large buoyant sphere in isotropic turbulence. We introduce an efficient method to simultaneously determine the position and (absolute)~orientation of a spherical body from visual observation. The method employs a minimization algorithm to obtain the orientation from the 2D projection of a specific pattern drawn onto the surface of the sphere. This has the advantages that it does not require a database of reference images, is easily scalable using parallel processing, and enables accurate absolute orientation reference. Analysis of the sphere's translational dynamics reveals clear differences between the streamwise and transverse directions. The translational auto-correlations and PDFs  provide evidence for periodicity in the particle's dynamics even under turbulent conditions. The angular autocorrelations show weak periodicity. The angular accelerations  exhibit wide tails, however without a directional dependence.}


\section{Introduction}

Particles suspensions in turbulent flows can be found in a wide range of natural and industrial settings. The behavior of these particles depends on several parameters including their size and density ratio, the particle Reynolds number, the particle Stokes number and the Taylor Reynolds number of the carrier turbulent flow. Understanding how the dynamics of particles is influenced by these parameters is crucial to make predictions on global phenomena of interest such as pollutant transport, cloud formation, and mixing in industrial processes~(\citet{toschi2009lagrangian}).

In many natural settings, particles can have a large size, and their density can be different from the carrier fluid.
Theoretical studies often model such objects as passive, finite-sized particles advected by the flow~(\citet{maxey1983equation,calzavarini2009acceleration,biferale2005particle}). 
\textcolor{black}{These are applicable in the limit of vanishing particle and shear Reynolds numbers.} However, in most practical situations, the density-mismatch of particles with the fluid results in finite drift velocities, which leads to finite particle Reynolds numbers~(\citet{jimenez1997oceanic}, \citet{mackenzie1993wind}, \citet{skyllingstad1999upper}, \citet{mathai2015wake}~etc). In such situations, particle dynamics can be strongly influenced by the unsteady wake-induced forces. 
For instance, a large buoyant sphere freely rising through quiescent fluid displays rich variability in translational dynamics~(\citet{ern2012wake, horowitz2008critical, horowitz2010effect}). The forcing responsible for such varied dynamics is linked to the vorticity shed in the wake of the particle~(\citet{achenbach1974vortex}, \citet{govardhan2005vortex}). Since these forces may not act along the geometric centre of the particle, it is possible that these could as well induce torques on the body. Little is known about the resulting translational and rotational dynamics, particularly for the case of buoyant particles in turbulence, which motivates us to develop a reliable measurement technique for studying these issues.

In three dimensions, an object's location and orientation can be fully described by six independent variables. Many physical experiments rely on image analysis to obtain these parameters from experimental data~(\citet{bovik2010handbook, meyer2013rotational}). Most of these systems capture translation and retrieve orientation from relative motion of translating nodes~(\citet{klein2013simultaneous}). \textcolor{black}{In the more elementary forms, the translation of nodes could be used to determine the velocities of the particles. However, these methods were not accurate enough when higher derivatives of orientation were to be determined. \citet{zimmermann2011tracking} introduced a method based on the identification of possible orientation candidates at each time step using projections of a pattern painted on the surface of a sphere.} 
\textcolor{black}{They found surprisingly intermittent behavior in the acceleration statistics of a neutrally buoyant sphere of diameter of the order of the integral scale.}

\textcolor{black}{In the present work, we introduce buoyancy to the problem of spherical particle dynamics in turbulence. We recover the translation and rotation as a function of time.} 
\textcolor{black}{The core of the method, which is to compare experimental images to synthetic ones is the same as proposed by~\citet{zimmermann2011tracking}. The novelty here lies in the way the pattern is generated (both physically on the surface of the particle and numerically for the comparison with actual images to match the orientation). Hence, the synthetic images for any given orientation are analytically known and do not need to be determined from static images. Furthermore, the method is easily scalable using parallel processing, which enables an accurate absolute orientation reference.} These aspects are elucidated in section~\ref{sec:method}. In addition, we describe a smoothing spline based roughness limiting technique that enables accurate representation of the higher derivatives of experimental data. Finally, in section~\ref{sec:res_disc}, we present the main results of our investigation on translational and rotational dynamics.

\section{Experimental Setup}

 \begin{figure}[!htbp]
 \begin{center}
 \includegraphics[width=0.44\textwidth]{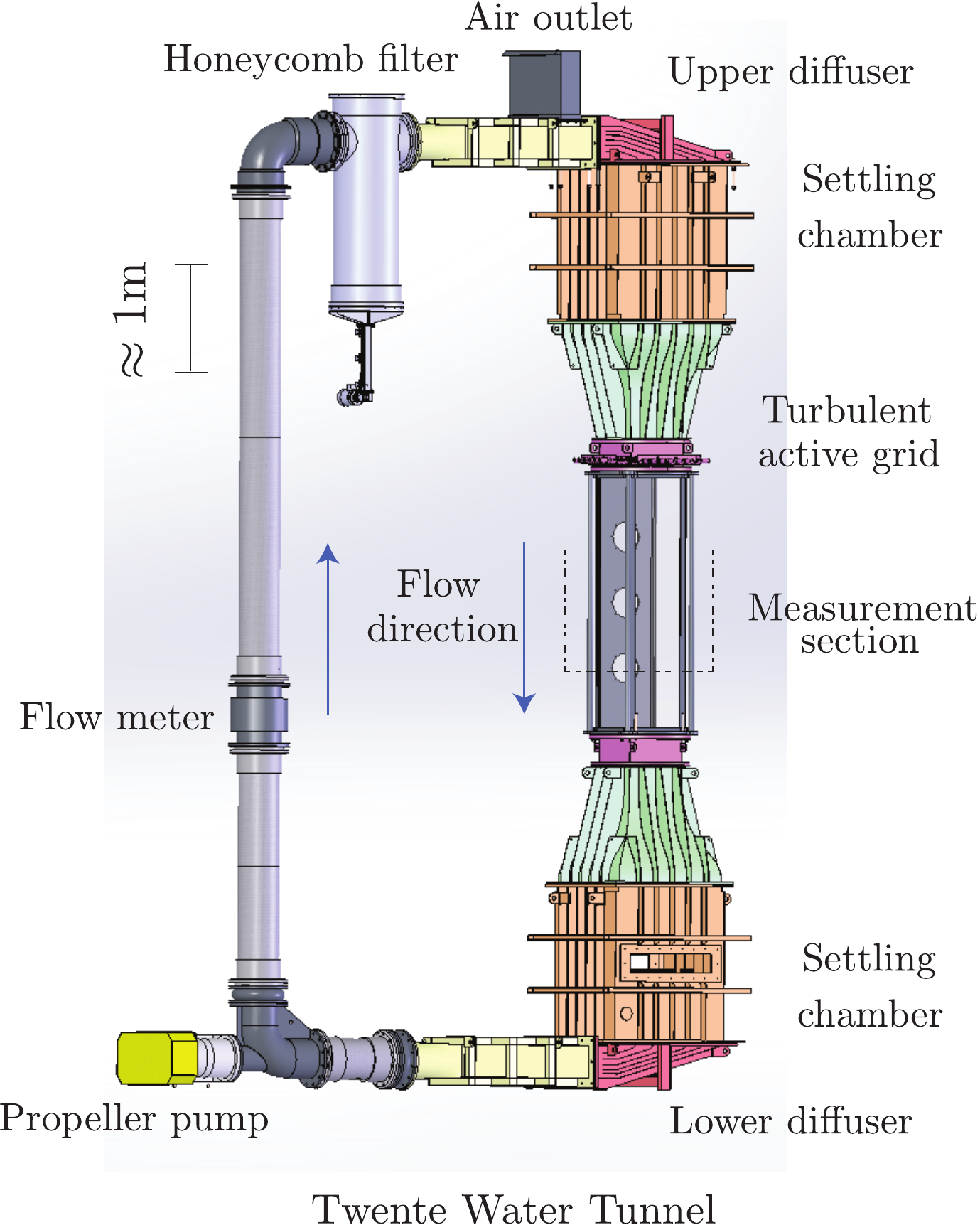}
  \caption{Sketch of the Twente Water Tunnel facility used for the measurements reported in the present study.}
\label{TWT}
\end{center}
\end{figure}

The experiments were conducted in the Twente Water Tunnel facility (TWT), designed to study particle-laden flows~(see figure~\ref{TWT} and \citet{rensen2005effect}). The measurement section has dimensions $0.45~\times~0.45~\times~2$~m$^3$, with three glass walls providing optical access to perform particle tracking experiments.
The setup houses an active grid above the measurement section, consisting of 24 independently rotating motors, which produce nearly homogeneous and isotropic turbulence with Re$_{\lambda} $ up to 300 in the downstream section of the water tunnel. The flow in the measurement section was characterized using a cylindrical hot film probe (Dantec 55R11) by following the same methodology as reported in \citet{mercado2012lagrangian}. The experiment reported here was performed at  $Re_{\lambda} \approx 300$. The dissipation rate $\epsilon=505\times10^{-6}$~m$^2$/s$^3$, and the dissipation length and times scales were approximately 211~$\mu$m and 44~ms respectively. The sphere used in the study has a diameter, $d_p$~=~25~mm, with an effective mass ratio, $ m^*\approx0.82$, where m* is the ratio of mass of sphere to mass of the sphere's volume of water~(see ~\citet{govardhan2005vortex}). The mass ratio was chosen such that the mean rise velocity of the sphere matched the mean downward flow velocity in the measurement section. This was necessary for obtaining sufficiently long sphere trajectories for well converged Lagrangian statistics. The spheres were designed as spherical shells with the \textcolor{black}{MP300 resin, with a bulk density~$\approx$~1089~kg/m$^3$.} \textcolor{black}{The angular inertia of the sphere was comparable to that of a homogeneous sphere of water, as was the case in the study by \citet{zimmermann2011tracking}. The spherical shells were made using a 3D printing technique, and the surface roughness was within 50~microns. The printing resolution depended on two factors. First, the resolution of the 3D printer used for making the stencil. In the present case, we could produce complex stencil designs with dimensions as small as a millimetre on a 25~mm sphere. The second limiting factor was the painting procedure itself. Once the stencil was fixed on the sphere, the pattern had to be spray painted. Here again, it was practical to spray paint patterns of 2~mm smaller dimension.}

\begin{figure}[!htbp]
 \begin{center}
 \includegraphics[width=0.45\textwidth]{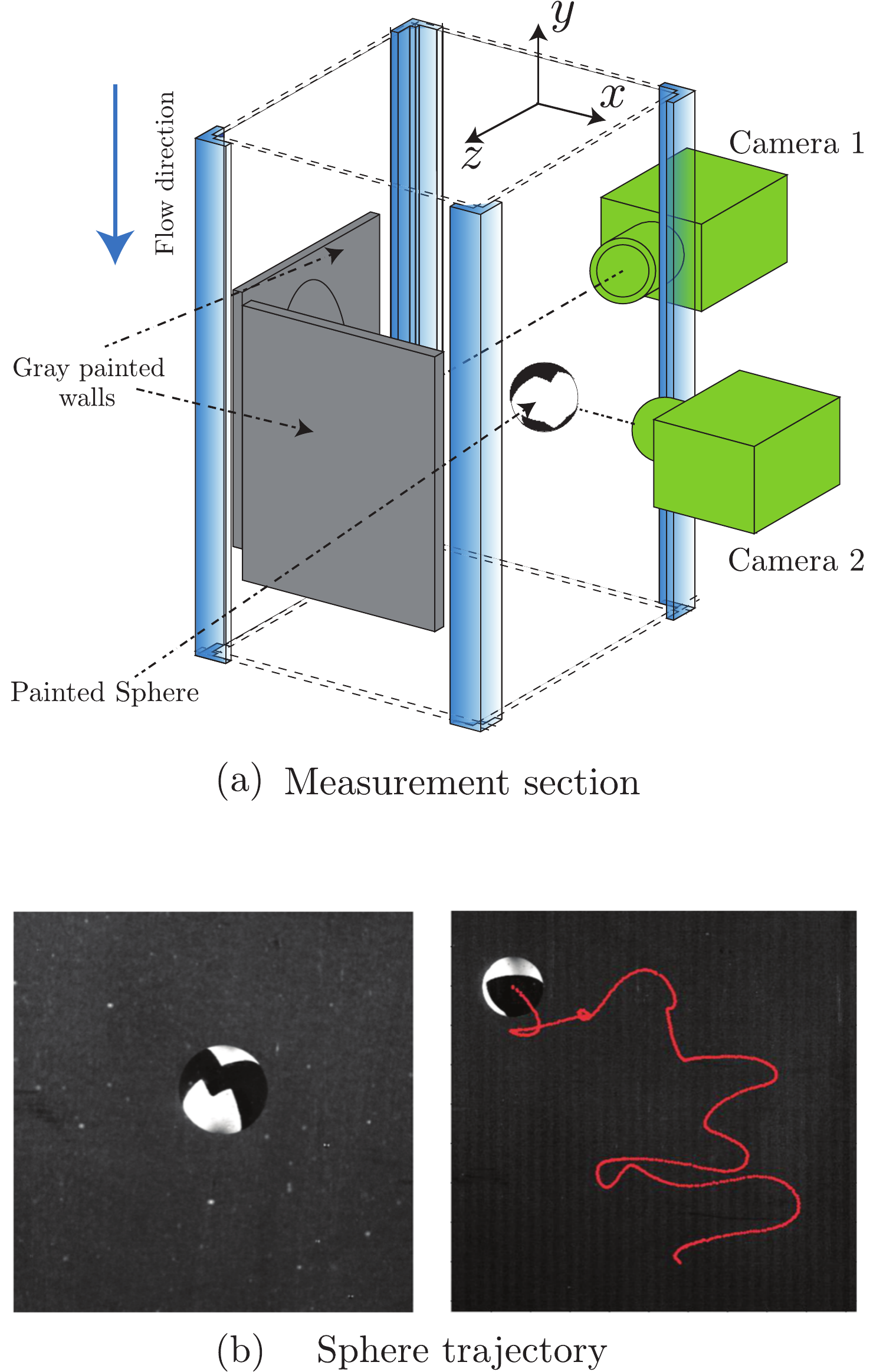}
  \caption{(a)~Measurement section of the Twente water tunnel with orthogonal camera experimental arrangement. A painted sphere is shown, which is viewed by both cameras.(b)~The sphere and its trajectory as recorded by one of the cameras.}
\label{Setup}
\end{center}
\end{figure}

The recordings were made with two Photron PCI-1024 high-speed cameras at 500 fps and megapixel resolution~(see figure~\ref{Setup}(a) \& (b)). The cameras were positioned at a 90$^{\circ}$ angle between them and focused at the center of the test section on a 150$\times$150 mm$^2$ area, which resulted in a spatial resolution of 150~$\mu$m/pixel. \textcolor{black}{The images showed that perspective effects were negligible. This was done by placing a sphere of known orientation at the corners of the field of view. The retrieved orientation varied less than 3$^\circ$.} 
The measurement volume was lit by eight 20~W LED lamps from the sides. The flow velocity was tuned to ensure that the spheres stayed in the viewing window for considerable duration.

\section{Method}
\label{sec:method}
\begin{figure} 
 \begin{center}
\includegraphics[width=0.45 \textwidth]{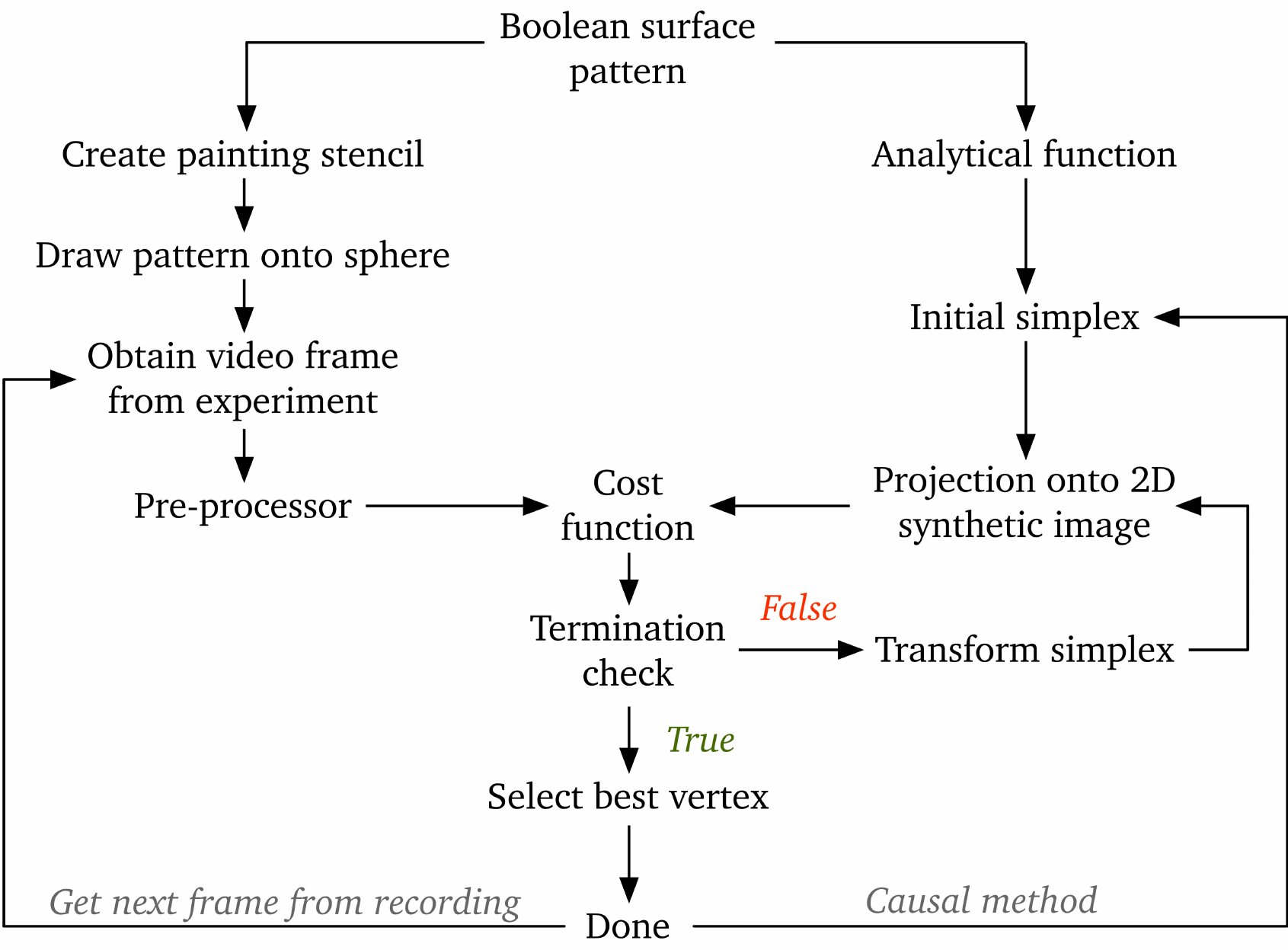}
  \caption{Orientation detection method flowchart}
\label{method_flowchart}
\end{center}
\end{figure}

\textcolor{black}{}The position and orientation of the sphere are determined using image analysis methods. The background was painted gray for contrast between the black and white pattern on the sphere. The sphere is separated from the background by subtracting the absolute difference in intensity from the background image. \textcolor{black}{Subtracting the absolute difference yields a full dark circle with a brighter background. The sphere is then detected using the Circular Hough transform~(CHT) technique and the centroid of the detected circle gives its position. The accuracy of CHT varied with angle of view. Therefore, choosing a good pattern is one of the most important steps. We found that a pattern that contained almost equal fraction of dark and bright areas for most angles of view improved the circle detection accuracy. Additionally, we used the outputs from previous frames to improve and speed up the detection process.} The CHT also returns the detected circle diameter $D_c$ in pixels from the image. Since the sphere diameter~(25~mm) is known, we use a calibration~(25/$D_c$)~mm/pixels for the recorded images. This ensures a correction for the minor magnification changes due to the sphere moving forward or backward in the measurement section.

\textcolor{black}{There are several approaches to obtaining the three dimensional trajectory of a particle. One approach is to reconstruct the spatial positions using multiple cameras~(\citet{ouellette2006quantitative}). Precise spatial reconstruction is needed when there are multiple particles in the image and  identifying and separating particles can be a challenge. In the present work, there are only a few particles in the measurement volume. We use a different method using two orthogonal cameras to obtain the three dimensional trajectory. The camera arrangement is such that the magnification and field of view are comparable between the two cameras. The camera field of view covered the full width of the water tunnel. We first obtain two dimensional trajectories of particles from both cameras. The redundant data corresponding to the vertical motion for two cameras is used to compare individual trajectories. In order to avoid ambiguities, we cross-correlate the vertical acceleration time series from the two cameras. A match is said to be found when the cross correlation of two acceleration time series yields a coefficient greater than 98 percent, even for long trajectories such as the one shown in figure~\ref{Setup}(b). This yields a simple yet robust three dimensional track of the particles without needing to resort to complex spatial reconstruction algorithms.}

\begin{figure} 
 \begin{center}
\includegraphics[width=0.41 \textwidth]{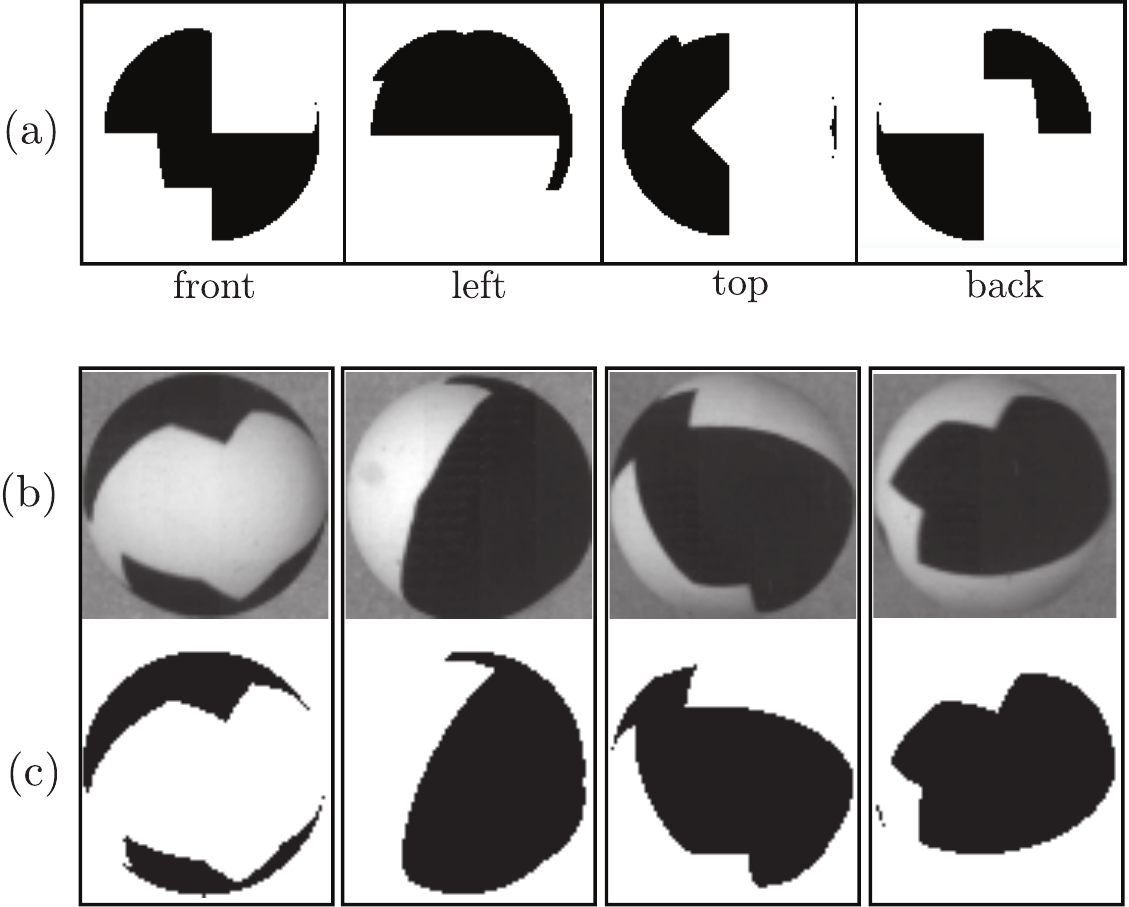}
  \caption{(a)~Several viewpoints of the synthetic pattern. Here, the synthetic projection is defined in the axis-angle convention as front: $\rho=(1, 0, 0, 0^{\circ}$), left: $\rho=(0, 0, 1, -90^{\circ}$), top: $\rho = (0,1,0,-90^{\circ}$) and back: $\rho = (0,0,1,180^{\circ}$). (b)~Experimental images of the sphere with the painted pattern for arbitrary orientations. (c)~Synthetic image equivalents as found by the orientation detection program corresponding to the experimental images in (b).} 
\label{SynthProjection}
\end{center}
\end{figure}

Obtaining the rotation of the sphere is a more complex task compared to position tracking. Hence, the remainder of this section will be used to describe the method that is used to obtain the rotation. The method to determine the rotation can be divided into four parts. Initially, a suitable boolean surface pattern is created. This pattern can be described as a piecewise constant analytic function $F \left( \theta, \phi \right)$ such that, given a coordinate on the surface of the sphere, the function returns either a zero or a one depending on the color of its corresponding infinitesimally small surface element. Here, $\theta$ and $\phi$ are the azimuthal and the polar angles respectively, and $F \left( \theta, \phi \right)$ is radius independent. Second, the pattern is drawn onto a physical sphere. This is realized using a 3D-printed painting stencil and an airbrush system. It is imperative that $F \left( \theta, \phi \right)$ is painted as accurate as possible to decrease the introduced error in this step. Third, a synthetic 2D image is constructed from a projection of the surface of the sphere onto a plane. \textcolor{black}{The synthetic image for any given orientation is analytically known and does not need to be determined from static images.} The projection is a function of the angle of rotation of the sphere and can be conceptually understood as the analog of the projection of the physical sphere on the recording camera. Finally, the rotated and projected synthetic pattern is compared to an image of the physical sphere. \textcolor{black}{The minimizing function can take different forms. For instance, a cross-correlation function between the synthetic pattern and the camera image could be used to find the best match. Alternately, a suitable cost function may be used to search for a match. Here we use a cost function, defined as the sum of the absolute difference between the binarized image pixels and the corresponding pixels in the synthetic image.} The orientation for which this comparison yields the best match is then determined using a Nelder-Mead minimization algorithm. These steps are illustrated in a flowchart in figure~\ref{method_flowchart}. The Fortran90 code and the stl-file of the stencil used for painting the spheres have been included as supplemental material.

\begin{figure} [!htbp]
 \begin{center}
\includegraphics[width=0.4\textwidth]{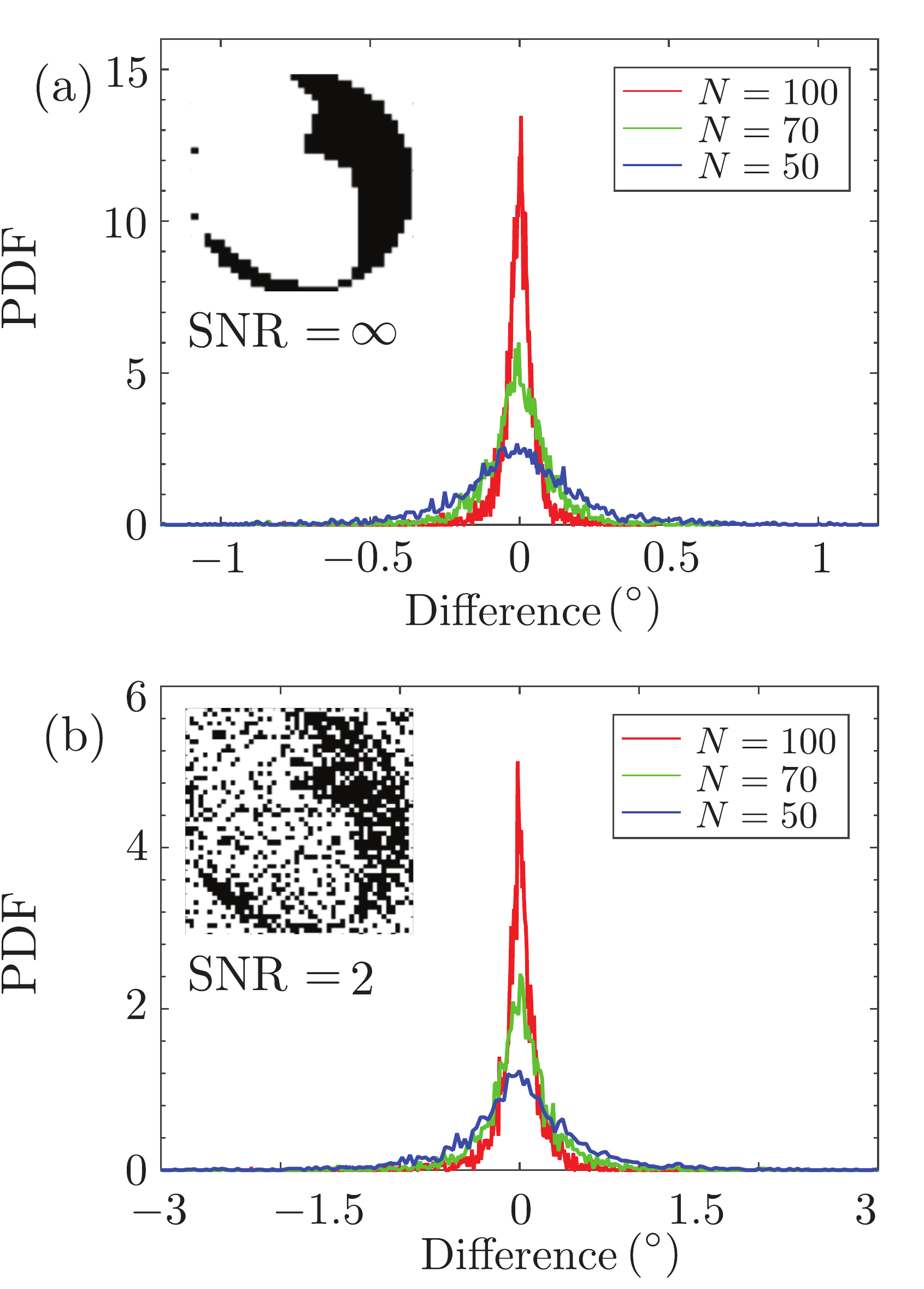}
  \caption{PDFs of the difference between actual orientation and orientation as found by the orientation detection method for different Signal-to-noise ratios (SNR). \textcolor{black}{a)~SNR = $\infty$ and b)~SNR = 2. Here $N$ is the width of the sphere in the image in pixels. $N$ takes on values 50, 70 ad 100 in the three cases shown. With increasing resolution $N$, the width of the PDF decreases.}}
\label{PDF_SNR}
\end{center}
\end{figure}

\textcolor{black}{The choice of painted pattern is an important step in the method of orientation detection. A necessary condition is that the projection of $F \left( \theta, \phi \right)$ onto a plane must be unique for each orientation of the sphere. Additionally, it is desirable that the pattern contains a minimum of edges and corners. The latter criterion is essential for fast convergence of the algorithm to the global minimum. Here we use the axis-angle method to describe the orientation considering its straightforward and singularity-free definition. It allows for smooth and continuous rotation from any orientation including around the Euler-angle singularities as it does not suffer from gimbal lock problems. }

\textcolor{black}{The pattern $F \left( \theta, \phi \right)$ represents a simply connected region, even then, some local minima may still arise due to the two-tone color limit. When the orientation is unknown, for instance in the first frame of a movie, a comparison of several initial orientation estimates solves this local minima problem. This yields the global minima. However, every initial estimate reduces the performance. Fortunately, given a sufficiently high ratio of frame rate over rotation rate, any frame in a sequence may use the outcome of the previous frame as initial estimate, reducing the number of initial estimates required to just one. This \emph{causal} method is the preferred method for analyzing large sequential datasets.} The computational performance of the method is approximately 30 frames per second on a contemporary computer, suggesting that processing of large datasets is straightforward.

%

\begin{figure}
 \begin{center}
 \includegraphics[width=0.42 \textwidth]{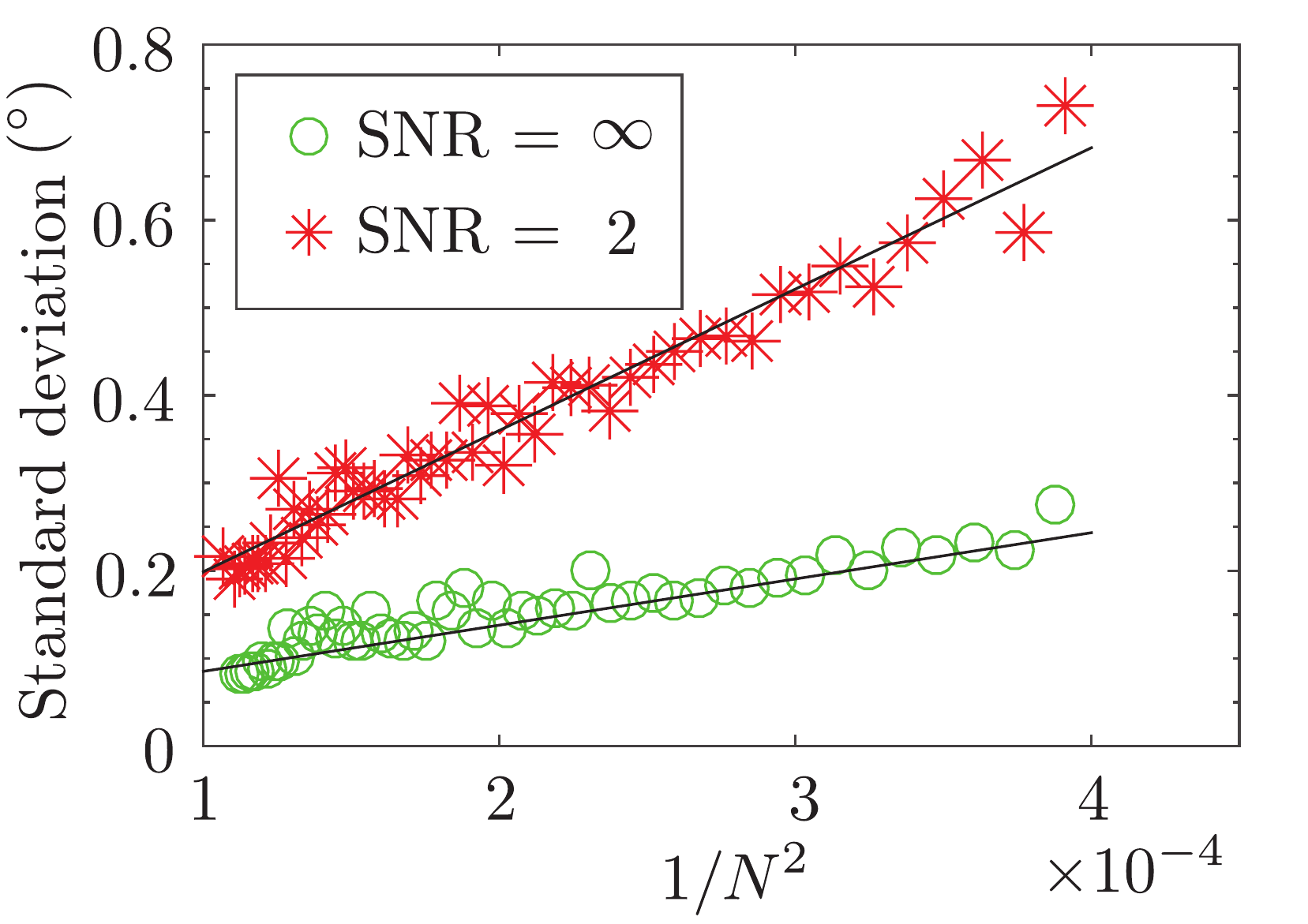}
  \caption{Numerical error scaling R$^2$ of linear fits are 0.87 and 0.96 for the SNR = $\infty$ and SNR = 2 respectively.}
\label{err_sc1}
\end{center}
\end{figure}

To determine the numerical accuracy of the method, several sets of 1024 projections are created using pseudo-random generated orientations. These synthetic images are used as input to our algorithm. A probability density function of the difference between the actual orientation and the orientation as found by the algorithm is shown in figures \ref{PDF_SNR}(a)~\&~(b), and the corresponding scaling of the standard deviation as function of the image width $N$~(in pixels) is shown in figure \ref{err_sc1}. The standard-deviation of accuracy scales as $\mathcal{O}(N^{-2})$, and assuming a Gaussian error distribution, this means that less than 1\% of the measured data showed an error larger than 1 degree for $N \geq 50$. Also, figure \ref{err_sc1} shows that noise decreases the accuracy to some extent but does not affect the reliability of the algorithm. \textcolor{black}{In the experiment, most image artifacts arise due to shadows and glare, which may be reduced by using diffuse light-sources and a matte-paint finish on the surface of the sphere.}

\begin{figure*} [!htbp]
 \begin{center}
 \includegraphics[width=0.9 \textwidth]{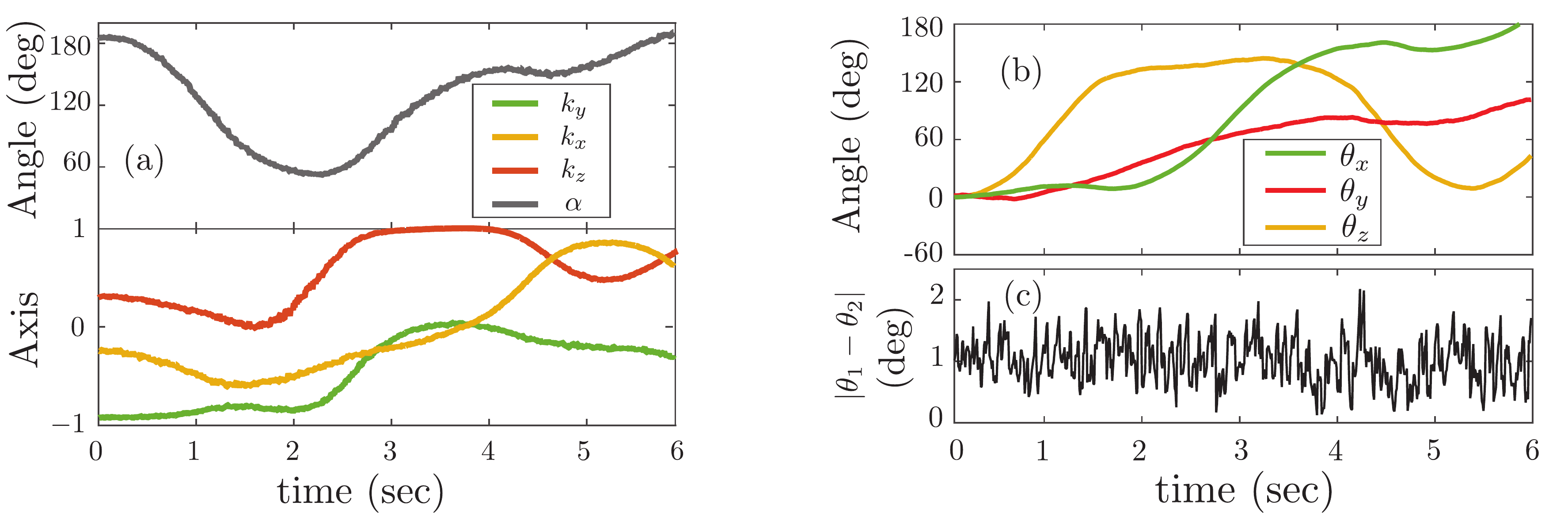}
  \caption{(a)~Axis-angle output obtained for a long trajectory of a buoyant sphere in a turbulent flow. (b)~Orientation~($\theta_x, \theta_y, \theta_z$) obtained from Camera~1. (c) Absolute difference between independent orientation measurements from Camera~1 and Camera~2. The maximum deviation is within 2.5~degrees. Mean of the difference is within 0.2~degrees.}
\label{orient_track}
\end{center}
\end{figure*}

\textcolor{black}{The axis-angle output, $\rho \equiv (k_x,k_y, k_z,\alpha)$, is defined with respect to a reference orientation in the camera coordinate system~(see leftmost image in figure~\ref{SynthProjection}~(a)). A typical output obtained from a buoyant sphere in a turbulent flow is given in figure~\ref{orient_track}(a). This output has little physical significance. Quantities of greater relevance to a rotating sphere are its rotational kinetic energy $I \omega^2$ and the net torque exerted on it by the surrounding fluid $I \alpha$, where $I$ is the moment of inertia of the sphere, and $\omega$ and $\alpha$ are the angular velocity and acceleration respectively. 
We adopt the following approach to compute $\omega$ and $\alpha$ in the lab coordinate system. Firstly, a high framing rate is used for accurate estimation of the time derivatives. In the present case, a frame rate of 500~fps ensures about 30 recordings in one Kolmogorov time. Therefore, within the inter-frame time interval $\Delta t = {1}/{500}$~sec, the particle's angular velocity may be assumed constant.}

\textcolor{black}{According to Euler's theorem, the angular velocity is linked to the inter-frame axis-angle, $\rho_{\Delta}~\equiv~(k_{\Delta x}, k_{\Delta y}, k_{\Delta z}, \Delta\alpha)$, by the relation~
\begin{equation}
\vec{\omega} (t) = \frac{\Delta \alpha}{\Delta t} (k_{\Delta x} \hat{i}  + k_{\Delta y} \hat{j} + k_{\Delta z} \hat{k} )
\label{angularvelocity}
\end{equation}}

\textcolor{black}{Now $\rho_{\Delta}$ may be obtained from the reference orientation based axis-angle output~($\rho$) through a few transformations. Firstly, the axis angle output is converted to rotation matrix using Rodrigues formula.  Let $[R_{0,i}]$ and $[R_{0,(i+1)}]$ represent the rotation matrices corresponding to rotations from the reference orientation to the orientations in $i^{th}$ and $(i+1)^{th}$ images respectively. Then $[R_{0,i}] \times [R_{i,(i+1)}] = [R_{0,(i+1)}$] and consequently, the inter-frame rotation matrix $[R_{i,(i+1)}]=[R_{0,i}]^T \times[R_{0,(i+1)}]$. This follows from the identity that $[R]^{-1} = [R]^T$ for rotation matrices.  From $[R_{i,(i+1)}]$, $\rho_\Delta \equiv \rho_{i,(i+1)}$ may be obtained, and eq.~\ref{angularvelocity} gives the angular velocity $\vec{\omega} (t)$ in the lab coordinate system.}

%

\textcolor{black}{The accuracy of the detection in a real experiment needs to be verified by other methods. We use a two-camera arrangement for this. If the orientations determined by the two cameras are comparable, then the method can be regarded accurate. Figure~\ref{orient_track}(b) shows the three orientation angles obtained from Camera~1 for a long particle trajectory. The same sphere was captured by Camera~2, which was placed at a 90$^{\circ}$ angle with Camera~1.  The difference in orientation prediction is plotted in figure~\ref{orient_track}(c). The maximum deviation is within 2.5 degrees. These differences are due to experimental noise and may be filtered out in the data smoothing step, to be explained in the following section.}

\section{Evaluating higher derivatives from experimental data}

\begin{figure}
 \begin{center}
 \includegraphics[width=0.44 \textwidth]{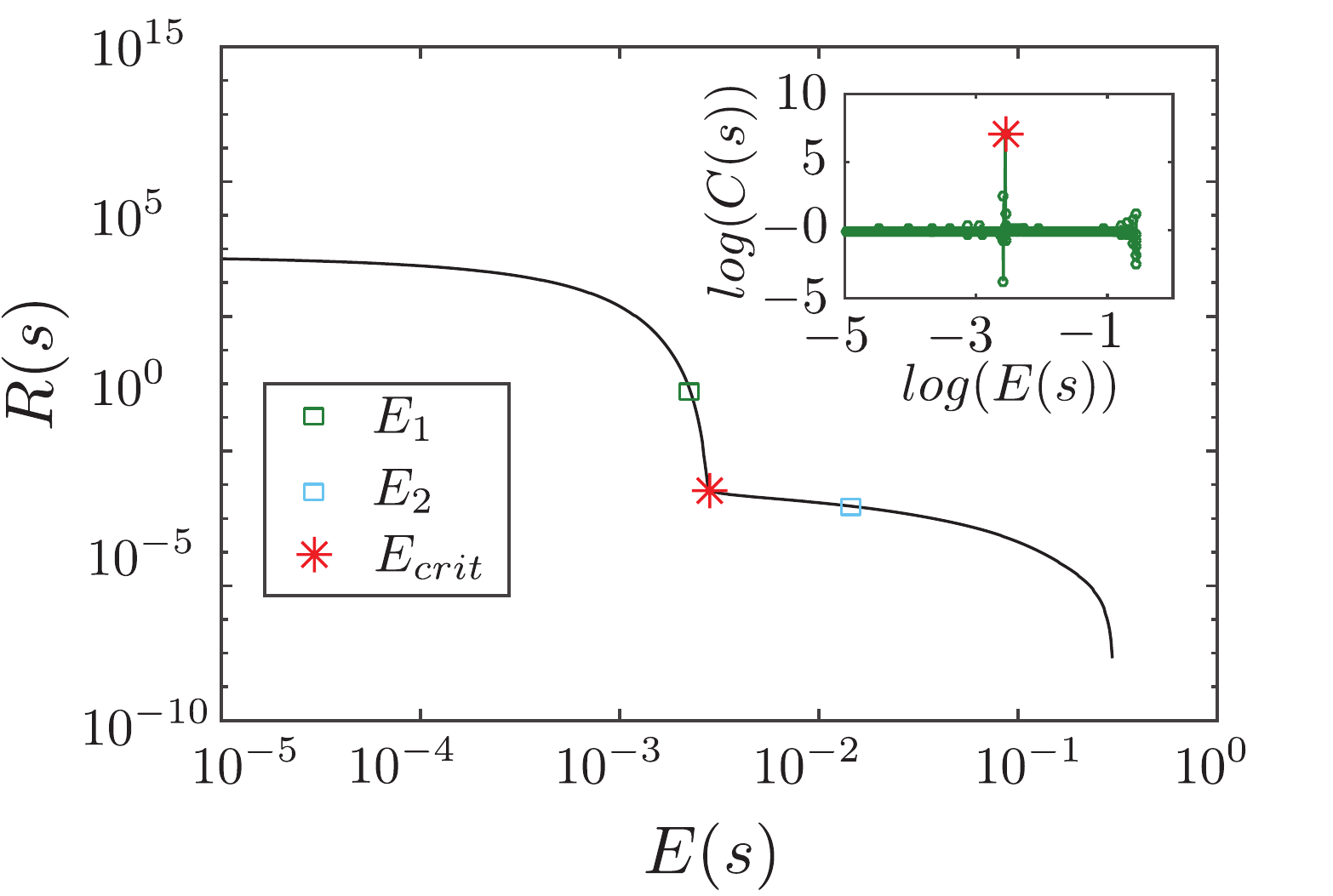}
  \caption{Roughness vs error frontier for a family of candidate smoothing splines. The kink marks the critical error tolerance of the optimal smoothing spline. \textcolor{black}{Inset shows $log(C(s))$ vs. $log(E(s))$. The optimal fit corresponds to the maximum curvature point indicated by the red star symbol.}}
\label{RvsE}
\end{center}
\end{figure}

\begin{figure*} [!htbp]
 \begin{center}
 \includegraphics[width=0.8 \textwidth]{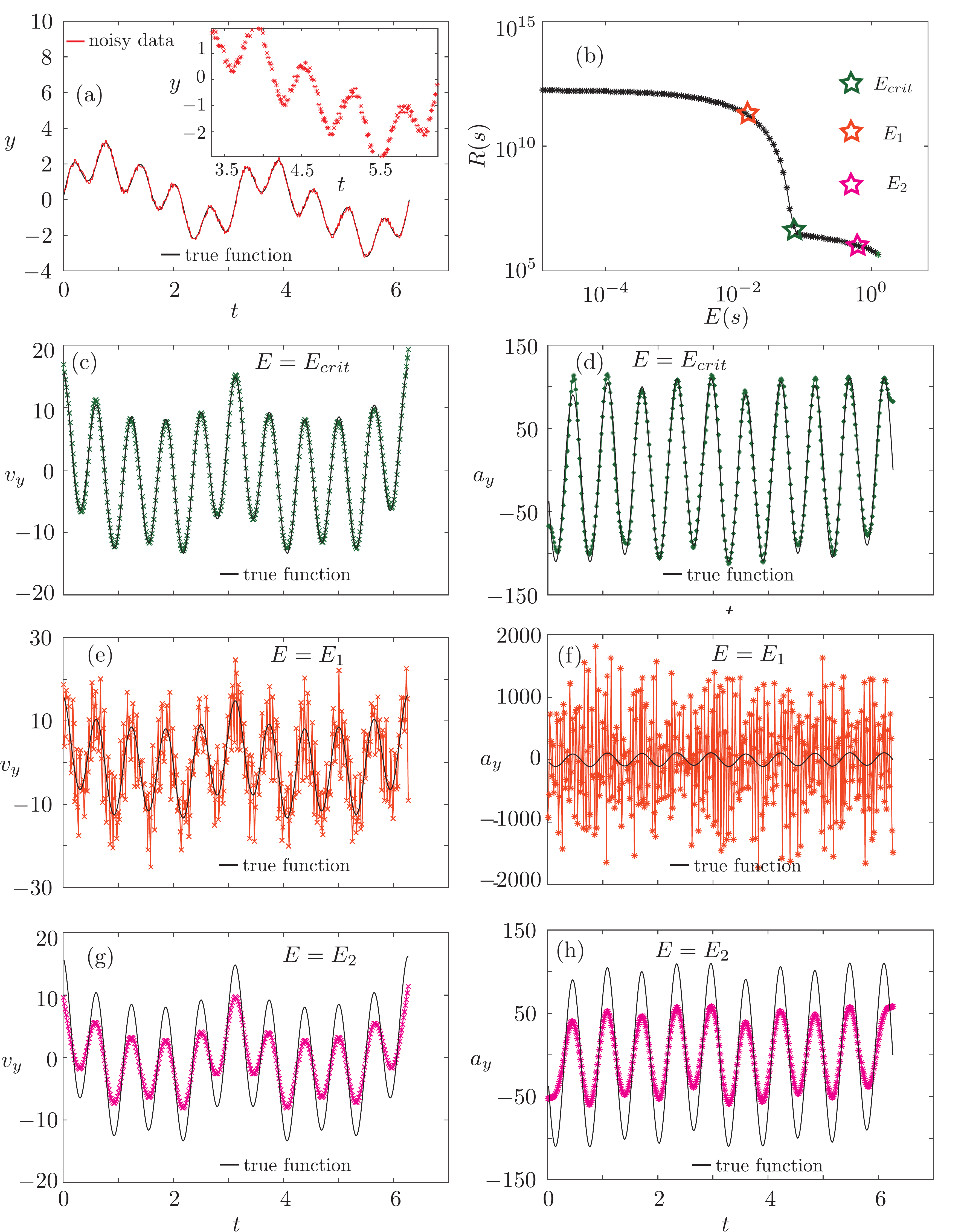}
  \caption{(a)~Analytical function and the noisy data with SNR $\approx$~4. Inset shows a zoom-in of the noisy data. (b)~Roughness vs error frontier for the noisy data. First and second derivatives (velocity and acceleration) estimated for (c) \& (d)~E = E$_{crit}$, (e) \& (f)~E = E$_{1}$, and (g) \& (h)~E = E$_{2}$. E =E$_{crit}$ gives an accurate estimate of even the second derivative, except at the endpoints.}
\label{Sample_spline}
\end{center}
\end{figure*}

Evaluating the forces and torques is an important step in understanding any dynamical system. One way to do this is by using accelerometers. However, this may not always be convenient since a direct measurement of acceleration usually requires attaching devices to the body. Moreover, the method is intrusive in nature and often leads to variations in the mass and center of gravity of the system (\citet{zimmermann2013characterizing}). In such situations, it may be suitable to numerically compute the acceleration from the second derivative of position of the body. This approach is the basis of our present study, where we determine both the position and orientation of a large buoyant sphere in turbulence using recordings from a high speed camera. 

Estimating the derivatives from experimentally determined position and orientation data is a nontrivial task. The difficulty arises because data obtained from experiments will have inherent noise due to measurement uncertainties. Historically, there have been two popular methods for smoothing particle trajectories in turbulent flows. One method involves fitting parts of the particle trajectory to a polynomial of second order or higher. \citet{voth2002measurement} used a second-order polynomial, \citet{luthi2005lagrangian} and \citet{mercado2012lagrangian} used a third-order polynomial. Other researchers have used a Gaussian kernel for smoothing (\citet{mordant2004experimental}, \citet{volk2011dynamics}). Both the mentioned methods employ piecewise discontinuous fitting of analytical functions to smooth out the noise. The effectiveness of these methods in filtering out the experimental noise depends on the fitting parameters chosen. In spite of the extensive literature on smoothing methods, there still prevails a general lack of consensus on the method of finding the optimal fitting parameter. Existing guidelines for the choice of the fitting parameters, such as those suggested by \citet{mordant2004experimental}, require knowledge of the smallest time scales in the flow apriori, and therefore are likely to introduce a bias into the analysis.

In this paper, we explore an alternate method using smoothing splines to filter out the experimental noise. The method is based roughly on the work by \citet{epps2010impulse}. While this was originally used to estimate the translational accelerations experienced by a water-entering object~(\citet{truscott2012unsteady,truscott2009cavity}), the method may be applied to particle trajectories in turbulence. \textcolor{black}{Consider a general set of experimental data $y(t)$ acquired at high temporal resolution. We use a smoothing spline based roughness limiting method to reduce the experimental noise in the data for obtaining higher derivatives. The function \textit{spaps} in Matlab is defined by two fitting parameters: the order of the smoothing spline and the error tolerance, $E(s) = \int_{t_1}^{t_N}~|y(t)-s(t)|^2dt$. We use a \textit{quintic} smoothing spline, which ensures that the second derivatives are properly represented. This leaves us with one fitting parameter, the error tolerance. We use a roughness estimate, $R(s)~=~\int_{t_1}^{t_N} |\frac{d^3s}{dt^3}|^2dt$, to scan for the most suitable fit to the experimental data. This is based on the assumption that the true function does not have very large changes in acceleration, which typically are due to noise in the data. In figure~\ref{RvsE}, we demonstrate the step to determine the optimal fit from experimental data. Increasing the error tolerance beyond $E_{crit}$ does not reduce the roughness of the curve. For error tolerances below $E_{crit}$, the roughness of the fits are increased significantly, indicating that the noise contained in the data is not properly removed. Therefore, the fit corresponding to the kink can be thought to best represent the true curve.}

\textcolor{black}{We test the sensitivity of the velocity and acceleration to the choice of the spline. For this, we use a sample analytical function defined as a combination of sine functions,  $y = a \sin{x} + b \sin{2x} + c \sin{4x} +d \sin{10x}$, where $a$, $b$, $c$ and $d$ were generated randomly in the [0~1] range. We introduce random noise to the signal with signal-noise-ratio $\approx 4$. The true function and noisy input are shown in figure~\ref{Sample_spline}(a). In figures~\ref{Sample_spline}(c)-(h), we compare the velocity and acceleration estimates for critical~(E$_{crit}$), sub-critical~(E$_1$) and super-critical~(E$_2$) error tolerances. The sub-critical error tolerance~(figure~\ref{Sample_spline}(f)) yields very high accelerations, while the super-critical error tolerance over-smoothes the curve~(figure~\ref{Sample_spline}(h)). Clearly, the acceleration estimate for critical error tolerance was obtained without prior knowledge of the timescales of the flow or the particle, and it compares fairly well with the analytical second derivative.}


\textcolor{black}{Another interesting feature of the present method is that the kink (figures~\ref{RvsE} \& \ref{Sample_spline}(b)) flattens out for experimental data with low signal-noise-ratios. Thus, the method also serves as a check for the quality of data, which can be difficult to estimate for standard smoothing methods. Alternately, the spline fit guess for the particle acceleration from a few representative tracks could be used to determined the optimal parameters for Gaussian Kernel smoothing~(\citet{ouellette2006probing}). This can speed-up the fitting process when sampling large datasets, while ensuring that the fitting windows are properly chosen.}

\section{Results and discussion}
\label{sec:res_disc}

\begin{figure}[!htbp]
 \begin{center}
 \includegraphics[width=0.395 \textwidth]{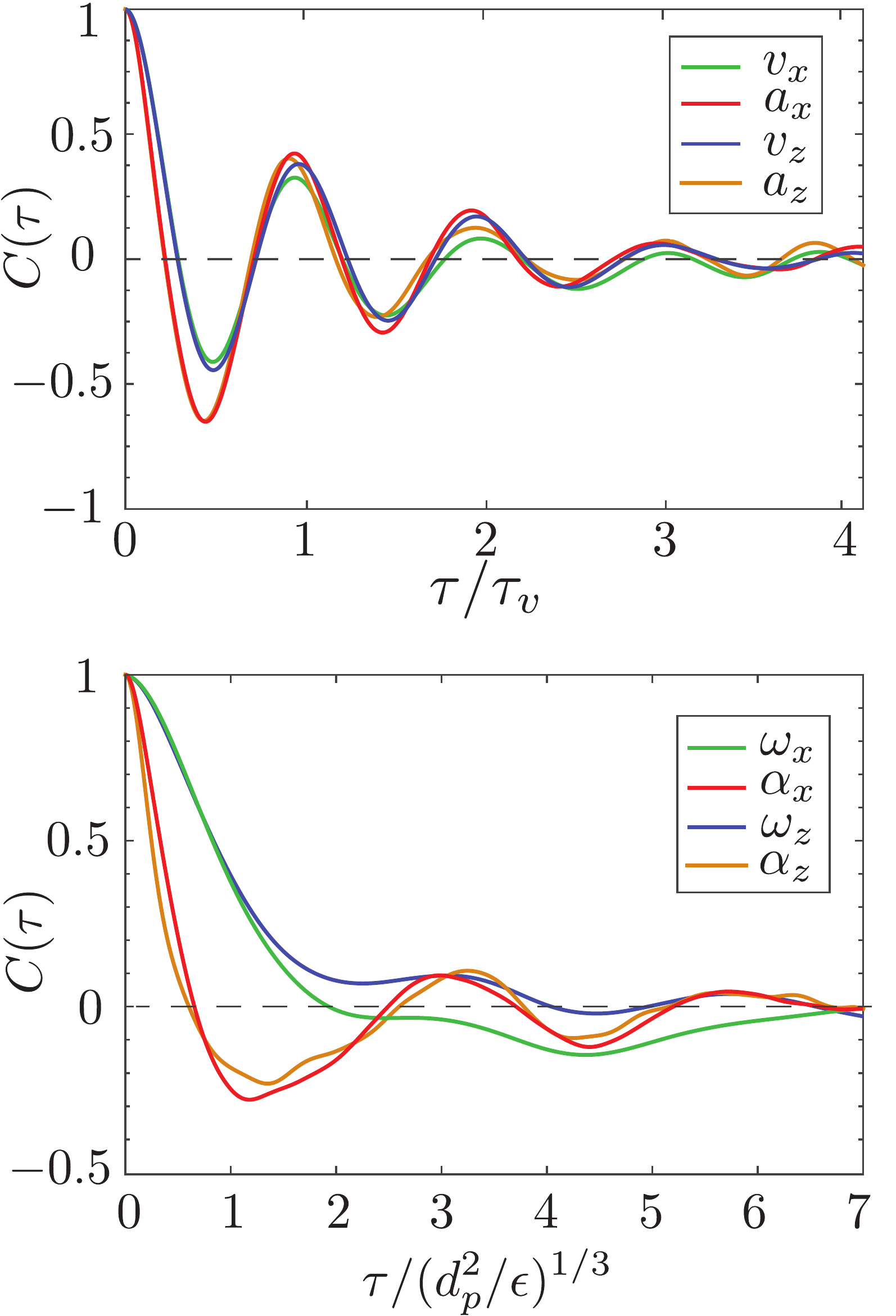}
  \caption{(a)~Lagrangian autocorrelation function of (a)~translational and (b)~rotational velocities and accelerations for the buoyant sphere in turbulence.}
\label{AutoCorr}
\end{center}
\end{figure}

\begin{figure*}
 \begin{center}
 \includegraphics[width= .8 \textwidth]{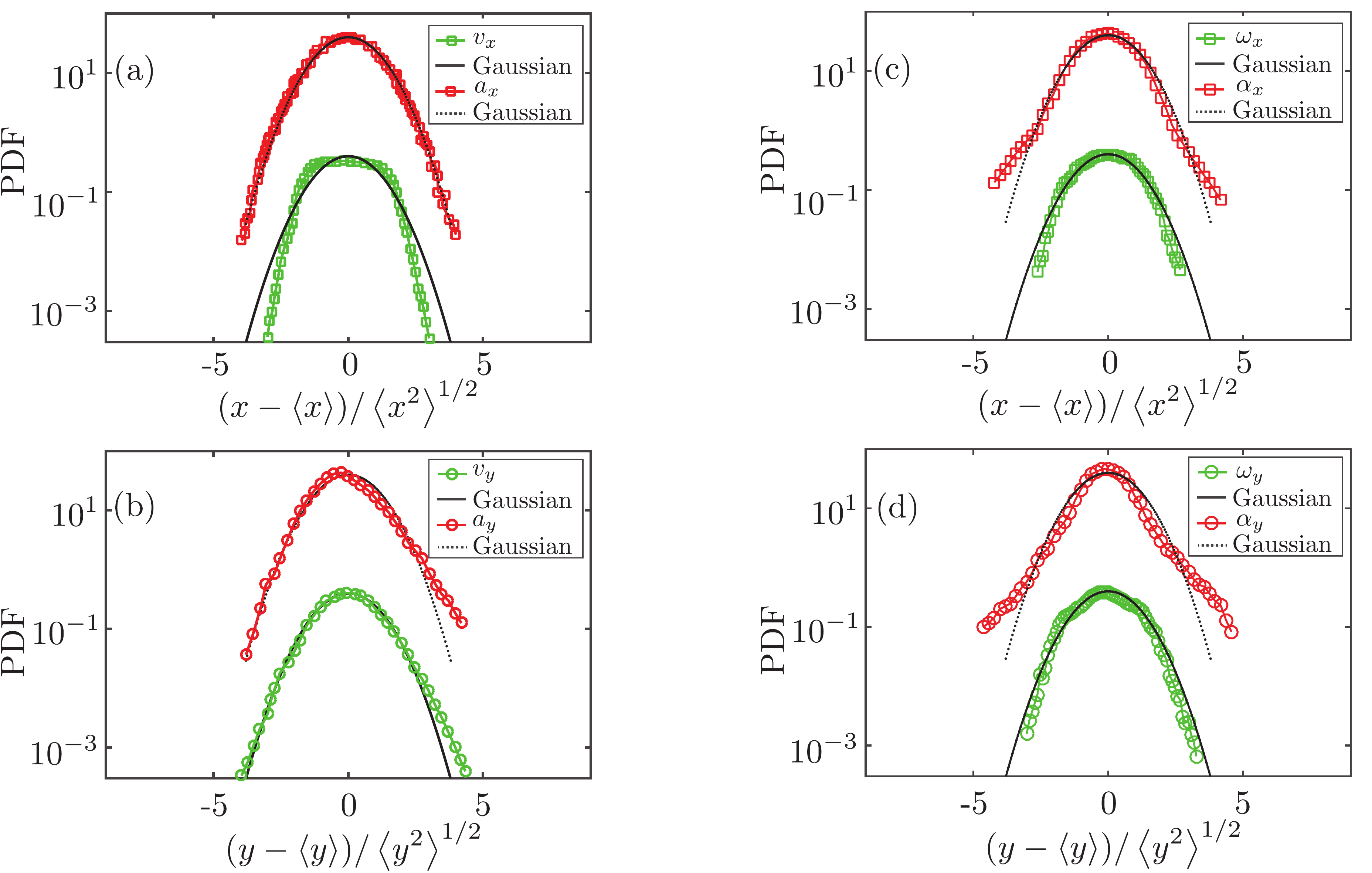}
  \caption{Velocity and acceleration PDFs in the (a)~transverse and (b)~streamwise directions for the buoyant sphere~(25 mm diameter). Angular velocity and angular acceleration PDFs in the (c)~transverse and (d)~streamwise directions for the sphere. The PDFs have been shifted vertically for clarity of viewing.}
\label{PDF_trans_XY}
\end{center}
\end{figure*}

\textcolor{black}{We present results on the translational and rotational motion of a large buoyant sphere in nearly homogeneous and isotropic turbulence.
We first address the question of how the particle's translational velocity and acceleration decorrelate in time. 
In Figure~\ref{AutoCorr}~(a), we plot the Lagrangian autocorrelation function for the horizontal~($x~\&~z$ directions) velocity and accelerations. \textcolor{black}{The particle response scale is fairly well predicted from the relation $\tau_v = d_p/(\textup{St}~\times~U_r)$, where St$-$Strouhal number~$\approx$~0.2, and $U_r$ is the measured mean rise velocity relative to the flow. Interestingly, both velocity and acceleration decorrelate in the same time $\sim 0.5$ sec, and they both display periodicity.}  This decorrelation behavior is fundamentally different from that of a passive finite-size particle in turbulence, for which the acceleration decorrelates in much shorter time than the velocity~(\citet{ishihara2009study}). \textcolor{black}{Therefore, in the present case, the dominant velocities and accelerations originate from vortices with the same timescale.}}

\textcolor{black}{Figure~\ref{AutoCorr}(b) shows the angular velocities and accelerations in the horizontal direction. \textcolor{black}{The time axis is normalised with the particle-sized eddy time scale, $\tau_d~=~(d_p^2/\epsilon)^{1/3} \sim$ 1 sec. The angular acceleration crosses the first minima in approximately $1~\tau_d$, which is four times longer compared to the minima crossing time for translational acceleration. The angular velocity components decorrelate even slower than the angular acceleration, and they show only a weak periodicity. The angular velocities and translational accelerations are weakly correlated, with a correlation coefficient $\approx$ 0.05. Surprisingly, we do not find any preferential orientation of the angular velocity vector with the translational acceleration vector, as found by~\citet{zimmermann2011rotational}. Therefore, in the present case, particle rotation does not appear synchronized with its translational motion, and we find no strong evidence suggesting a lift force. More detailed studies are to be conducted to gain further insights into the underlying physics. In future, we also aim to look into the flow structure around the sphere along with its motion.}}


\textcolor{black}{In figure~\ref{PDF_trans_XY}(a) \& (b), we show the probability density functions~(PDFs) of the horizontal and vertical components of the velocity and acceleration. The horizontal velocity PDF shows a symmetric flat-head distribution with sub-Gaussian tails. The behaviour is notably different from the nearly Gaussian velocity PDFs found for neutrally buoyant particles~(\citet{mercado2012lagrangian}). \textcolor{black}{The flat top of the PDF may be explained from a typical time series of velocity of the sphere. We observe that the particle undergoes repeated cyclic motions, weakly disturbed by the carrier flow. The velocities in these cycles lie in the $\pm$ 1.5 $ \left <a^2 \right >^{1/2}$ range. More extreme accelerations are less frequent and contribute to the low probability tails of the PDFs.} The horizontal acceleration PDF is  Gaussian. The vertical velocity and accelerations have positively skewed distributions, clearly indicating a directional dependence. \textcolor{black}{This strong anisotropy is not inherent in the carrier turbulent flow. The water tunnel flow was reported to be nearly isotropic in earlier studies~(\citet{mercado2012lagrangian}). Therefore, the anisotropy here is expected to originate from the up-down asymmetry set up by the vortex-shedding downstream of the sphere.} The PDFs of the rotational quantities reveal a different story~(Figure~\ref{PDF_trans_XY}(c) \& (d)). Both horizontal and vertical angular velocities follow a nearly gaussian distribution. The angular acceleration PDFs show symmetric distributions but with wide tails compared to the translational acceleration PDFs. It may be noted that Re$_{\lambda}$ and particle size ratio $\Xi = d_p/\eta$ are comparable to the experiment of \citet{zimmermann2011tracking}. Thus, changing only the particle-fluid density ratio brings in these new effects, leading to observable differences in particle dynamics in turbulence. Future experiments will be aimed at tracking simultaneously the particle and the flow around it.}




\section{Conclusions}

\textcolor{black}{We have conducted experimental measurements of the dynamics of a large buoyant sphere in a turbulent flow and described the methods and procedures to track simultaneously its position and orientation. A fast, accurate and adaptable method that determines the absolute orientation of the sphere in 3D space  is described and validated. The standard deviation of error for this method is approximately 0.2 degrees. Distortion using salt  \& pepper noise with SNR$ = 2$ increases the standard deviation of error to 0.7 degrees, however without impacting the reliability of the method.}

\textcolor{black}{We followed the approach of using the experimentally determined data to calculate time derivatives (velocity and acceleration) of position and orientation. To this end, we employed a smoothing spline based roughness limiting technique, which enables an accurate estimation of higher derivatives. The method has the advantage that it yields an optimal fit that is continuous and differentiable.}

\textcolor{black}{Our results on the velocity and acceleration statistics reveal that buoyant spheres have very different dynamics from the well-explored class of neutrally buoyant particles in turbulence (\citet{zimmermann2011tracking}, \citet{toschi2009lagrangian}, and \citet{homann2010finite}). We detect the influence of trailing wake, resulting in periodicity in the Lagrangian auto-correlations and anisotropy in the translational dynamics. The rotational velocity and acceleration PDFs show wide tails, however without any observable skewness in the streamwise and transverse directions. The present measurements provide clues about how a buoyant particle interacts with a turbulent flow. The methods presented here open up a new direction in the exploration of particle dynamics in turbulence.}
\section{Acknowledgments}
We thank Jon~Brons, Fedde~Gaastra, and Jaap~ Nieland for their significant contributions to experiments and analysis. We are grateful to Bram~Verhaagen and Tobias Förtsch for help with the 3D printing of spheres, and Michiel~van~Limbeek for his guidance in using the painting equipment. Finally, we thank Detlef~Lohse and Andrea~Prosperetti for various fruitful discussions.


\begin{thebibliography}{32}
\providecommand{\natexlab}[1]{#1}
\providecommand{\url}[1]{\texttt{#1}}
\expandafter\ifx\csname urlstyle\endcsname\relax
  \providecommand{\doi}[1]{doi: #1}\else
  \providecommand{\doi}{doi: \begingroup \urlstyle{rm}\Url}\fi

\bibitem[Achenbach(1974)]{achenbach1974vortex}
Elmar Achenbach.
\newblock Vortex shedding from spheres.
\newblock \emph{Journal of Fluid Mechanics}, 62\penalty0 (02):\penalty0
  209--221, 1974.

\bibitem[Biferale et~al.(2005)Biferale, Boffetta, Celani, Lanotte, and
  Toschi]{biferale2005particle}
Luca Biferale, Guido Boffetta, Antonio Celani, Alessandra Lanotte, and Federico
  Toschi.
\newblock Particle trapping in three-dimensional fully developed turbulence.
\newblock \emph{Physics of Fluids (1994-present)}, 17\penalty0 (2):\penalty0
  021701, 2005.

\bibitem[Bovik(2010)]{bovik2010handbook}
Alan~C Bovik.
\newblock \emph{Handbook of image and video processing}.
\newblock Academic Press, 2010.

\bibitem[Calzavarini et~al.(2009)Calzavarini, Volk, Bourgoin, L{\'e}v{\^e}que,
  Pinton, and Toschi]{calzavarini2009acceleration}
Enrico Calzavarini, Romain Volk, Micka{\"e}l Bourgoin, Emmanuel
  L{\'e}v{\^e}que, J-F Pinton, and Federico Toschi.
\newblock Acceleration statistics of finite-sized particles in turbulent flow:
  the role of fax{\'e}n forces.
\newblock \emph{Journal of Fluid Mechanics}, 630:\penalty0 179--189, 2009.

\bibitem[Epps(2010)]{epps2010impulse}
Brenden~P Epps.
\newblock \emph{An impulse framework for hydrodynamic force analysis: fish
  propulsion, water entry of spheres, and marine propellers}.
\newblock PhD thesis, Massachusetts Institute of Technology, 2010.

\bibitem[Ern et~al.(2012)Ern, Risso, Fabre, and Magnaudet]{ern2012wake}
Patricia Ern, Fr{\'e}d{\'e}ric Risso, David Fabre, and Jacques Magnaudet.
\newblock Wake-induced oscillatory paths of bodies freely rising or falling in
  fluids.
\newblock \emph{Annual Review of Fluid Mechanics}, 44:\penalty0 97--121, 2012.

\bibitem[Govardhan and Williamson(2005)]{govardhan2005vortex}
RN~Govardhan and CHK Williamson.
\newblock Vortex-induced vibrations of a sphere.
\newblock \emph{Journal of Fluid Mechanics}, 531:\penalty0 11--47, 2005.

\bibitem[Homann and Bec(2010)]{homann2010finite}
Holger Homann and Jeremie Bec.
\newblock Finite-size effects in the dynamics of neutrally buoyant particles in
  turbulent flow.
\newblock \emph{Journal of Fluid Mechanics}, 651:\penalty0 81--91, 2010.

\bibitem[Horowitz and Williamson(2008)]{horowitz2008critical}
M~Horowitz and CHK Williamson.
\newblock Critical mass and a new periodic four-ring vortex wake mode for
  freely rising and falling spheres.
\newblock \emph{Physics of Fluids (1994-present)}, 20\penalty0 (10):\penalty0
  101701, 2008.

\bibitem[Horowitz and Williamson(2010)]{horowitz2010effect}
M~Horowitz and CHK Williamson.
\newblock The effect of reynolds number on the dynamics and wakes of freely
  rising and falling spheres.
\newblock \emph{Journal of Fluid Mechanics}, 651:\penalty0 251--294, 2010.

\bibitem[Ishihara et~al.(2009)Ishihara, Gotoh, and Kaneda]{ishihara2009study}
Takashi Ishihara, Toshiyuki Gotoh, and Yukio Kaneda.
\newblock Study of high-reynolds number isotropic turbulence by direct
  numerical simulation.
\newblock \emph{Annual Review of Fluid Mechanics}, 41:\penalty0 165--180, 2009.

\bibitem[Jimenez(1997)]{jimenez1997oceanic}
JAVIER Jimenez.
\newblock Oceanic turbulence at millimeter scales.
\newblock \emph{Scientia Marina}, 61:\penalty0 47--56, 1997.

\bibitem[Klein et~al.(2013)Klein, Gibert, B{\'e}rut, and
  Bodenschatz]{klein2013simultaneous}
Simon Klein, Mathieu Gibert, Antoine B{\'e}rut, and Eberhard Bodenschatz.
\newblock Simultaneous 3d measurement of the translation and rotation of
  finite-size particles and the flow field in a fully developed turbulent water
  flow.
\newblock \emph{Measurement Science and Technology}, 24\penalty0 (2):\penalty0
  024006, 2013.

\bibitem[L{\"u}thi et~al.(2005)L{\"u}thi, Tsinober, and
  Kinzelbach]{luthi2005lagrangian}
Beat L{\"u}thi, Arkady Tsinober, and Wolfgang Kinzelbach.
\newblock Lagrangian measurement of vorticity dynamics in turbulent flow.
\newblock \emph{Journal of Fluid mechanics}, 528:\penalty0 87--118, 2005.

\bibitem[MacKenzie and Leggett(1993)]{mackenzie1993wind}
BR~MacKenzie and WC~Leggett.
\newblock Wind-based models for estimating the dissipation rates of turbulent
  energy in aquatic environments: empirical comparisons.
\newblock \emph{Marine Ecology-Progress Series}, 94:\penalty0 207--207, 1993.

\bibitem[Mathai et~al.(2015)Mathai, Prakash, Brons, Sun, and
  Lohse]{mathai2015wake}
Varghese Mathai, Vivek~N. Prakash, Jon Brons, Chao Sun, and Detlef Lohse.
\newblock Wake-driven dynamics of finite-sized buoyant spheres in turbulence.
\newblock \emph{Phys. Rev. Lett.}, 115:\penalty0 124501, 2015.

\bibitem[Maxey and Riley(1983)]{maxey1983equation}
Martin~R Maxey and James~J Riley.
\newblock Equation of motion for a small rigid sphere in a nonuniform flow.
\newblock \emph{Physics of Fluids (1958-1988)}, 26\penalty0 (4):\penalty0
  883--889, 1983.

\bibitem[Mercado et~al.(2012)Mercado, Prakash, Tagawa, Sun, Lohse,
  et~al.]{mercado2012lagrangian}
Julian~Martinez Mercado, Vivek~N Prakash, Yoshiyuki Tagawa, Chao Sun, Detlef
  Lohse, et~al.
\newblock Lagrangian statistics of light particles in turbulence.
\newblock \emph{Physics of Fluids (1994-present)}, 24\penalty0 (5):\penalty0
  055106, 2012.

\bibitem[Meyer et~al.(2013)Meyer, Byron, and Variano]{meyer2013rotational}
Colin~R Meyer, Margaret~L Byron, and Evan~A Variano.
\newblock Rotational diffusion of particles in turbulence.
\newblock \emph{Limnology and Oceanography: Fluids and Environments},
  3\penalty0 (1):\penalty0 89--102, 2013.

\bibitem[Mordant et~al.(2004)Mordant, Crawford, and
  Bodenschatz]{mordant2004experimental}
Nicolas Mordant, Alice~M Crawford, and Eberhard Bodenschatz.
\newblock Experimental lagrangian acceleration probability density function
  measurement.
\newblock \emph{Physica D: Nonlinear Phenomena}, 193\penalty0 (1):\penalty0
  245--251, 2004.

\bibitem[Ouellette et~al.(2006)Ouellette, Xu, and
  Bodenschatz]{ouellette2006quantitative}
Nicholas~T Ouellette, Haitao Xu, and Eberhard Bodenschatz.
\newblock A quantitative study of three-dimensional lagrangian particle
  tracking algorithms.
\newblock \emph{Experiments in Fluids}, 40\penalty0 (2):\penalty0 301--313,
  2006.

\bibitem[Ouellette(2006)]{ouellette2006probing}
Nicholas~Testroet Ouellette.
\newblock \emph{Probing the statistical structure of turbulence with
  measurements of tracer particle tracks}.
\newblock Cornell University, 2006.

\bibitem[Rensen et~al.(2005)Rensen, Luther, and Lohse]{rensen2005effect}
Judith Rensen, Stefan Luther, and Detlef Lohse.
\newblock The effect of bubbles on developed turbulence.
\newblock \emph{Journal of Fluid Mechanics}, 538:\penalty0 153--187, 2005.

\bibitem[Skyllingstad et~al.(1999)Skyllingstad, Smyth, Moum, and
  Wijesekera]{skyllingstad1999upper}
Eric~D Skyllingstad, WD~Smyth, JN~Moum, and H~Wijesekera.
\newblock Upper-ocean turbulence during a westerly wind burst: A comparison of
  large-eddy simulation results and microstructure measurements.
\newblock \emph{Journal of physical oceanography}, 29\penalty0 (1):\penalty0
  5--28, 1999.

\bibitem[Toschi and Bodenschatz(2009)]{toschi2009lagrangian}
Federico Toschi and Eberhard Bodenschatz.
\newblock Lagrangian properties of particles in turbulence.
\newblock \emph{Annual Review of Fluid Mechanics}, 41:\penalty0 375--404, 2009.

\bibitem[Truscott et~al.(2012)Truscott, Epps, and Techet]{truscott2012unsteady}
Tadd~T Truscott, Brenden~P Epps, and Alexandra~H Techet.
\newblock Unsteady forces on spheres during free-surface water entry.
\newblock \emph{Journal of Fluid Mechanics}, 704:\penalty0 173--210, 2012.

\bibitem[Truscott(2009)]{truscott2009cavity}
Tadd~Trevor Truscott.
\newblock \emph{Cavity dynamics of water entry for spheres and ballistic
  projectiles}.
\newblock PhD thesis, Massachusetts Institute of Technology, 2009.

\bibitem[Volk et~al.(2011)Volk, Calzavarini, Leveque, and
  Pinton]{volk2011dynamics}
Romain Volk, Enrico Calzavarini, Emmanuel Leveque, and J-F Pinton.
\newblock Dynamics of inertial particles in a turbulent von k{\'a}rm{\'a}n
  flow.
\newblock \emph{Journal of Fluid Mechanics}, 668:\penalty0 223--235, 2011.

\bibitem[Voth et~al.(2002)Voth, la~Porta, Crawford, Alexander, and
  Bodenschatz]{voth2002measurement}
Greg~A Voth, Arthur la~Porta, Alice~M Crawford, Jim Alexander, and Eberhard
  Bodenschatz.
\newblock Measurement of particle accelerations in fully developed turbulence.
\newblock \emph{Journal of Fluid Mechanics}, 469:\penalty0 121--160, 2002.

\bibitem[Zimmermann et~al.(2011{\natexlab{a}})Zimmermann, Gasteuil, Bourgoin,
  Volk, Pumir, and Pinton]{zimmermann2011rotational}
Robert Zimmermann, Yoann Gasteuil, Mickael Bourgoin, Romain Volk, Alain Pumir,
  and Jean-Fran{\c{c}}ois Pinton.
\newblock Rotational intermittency and turbulence induced lift experienced by
  large particles in a turbulent flow.
\newblock \emph{Physical review letters}, 106\penalty0 (15):\penalty0 154501,
  2011{\natexlab{a}}.

\bibitem[Zimmermann et~al.(2011{\natexlab{b}})Zimmermann, Gasteuil, Bourgoin,
  Volk, Pumir, Pinton, et~al.]{zimmermann2011tracking}
Robert Zimmermann, Yoann Gasteuil, Mickael Bourgoin, Romain Volk, Alain Pumir,
  Jean-Fran{\c{c}}ois Pinton, et~al.
\newblock Tracking the dynamics of translation and absolute orientation of a
  sphere in a turbulent flow.
\newblock \emph{Review of Scientific Instruments}, 82\penalty0 (3):\penalty0
  033906, 2011{\natexlab{b}}.

\bibitem[Zimmermann et~al.(2013)Zimmermann, Fiabane, Gasteuil, Volk, and
  Pinton]{zimmermann2013characterizing}
Robert Zimmermann, Lionel Fiabane, Yoann Gasteuil, Romain Volk, and
  Jean-Fran{\c{c}}ois Pinton.
\newblock Characterizing flows with an instrumented particle measuring
  lagrangian accelerations.
\newblock \emph{New Journal of Physics}, 15\penalty0 (1):\penalty0 015018,
  2013.

\end{thebibliography}

\end{document}